\documentclass[11pt]{article}
\usepackage[margin=1.0in]{geometry}
\usepackage{amsmath,amssymb,graphicx}
\usepackage{hyperref}
\usepackage{slashed}

\usepackage[utf8]{inputenc}

\usepackage[font=small,labelfont=bf]{caption}
\usepackage{cite}
\usepackage{xcolor}
\usepackage{bm}
\usepackage{changepage}

\usepackage{atbegshi}
\AtBeginDocument{\AtBeginShipoutNext{\AtBeginShipoutDiscard}}% removes first blank page

\newcommand\snowmass{\begin{center}\rule[-0.2in]{\hsize}{0.01in}\\\rule{\hsize}{0.01in}\\
\vskip 0.1in Submitted to the Proceedings of the US Community Study\\
on the Future of Particle Physics (Snowmass 2021)\\
\rule{\hsize}{0.01in}\\\rule[+0.2in]{\hsize}{0.01in} \end{center}}

\title{\bf Snowmass White Paper: UV Constraints on IR Physics}
\author{Claudia de Rham${}^{1,2}$, Sandipan Kundu${}^3$, Matthew Reece${}^4$, Andrew J.~Tolley${}^{1,2}$ and Shuang-Yong Zhou${}^{5,6}$ \\
{\small ${}^1$Theoretical Physics, Blackett Laboratory, Imperial College London, SW7 2AZ London, U.K.}\\
{\small ${}^2$CERCA, Department of Physics, Case Western Reserve University, 10900 Euclid Ave, Cleveland, OH 44106, USA}\\
{\small ${}^3$Department of Physics and Astronomy, Johns Hopkins University, Maryland 21218, USA}\\
{\small ${}^4$Department of Physics, Harvard University, Cambridge, MA, 02138, USA}\\
{\small ${}^5$Interdisciplinary Center for Theoretical Study, University of Science and Technology of China, Hefei, Anhui 230026, China}\\
{\small ${}^6$Peng Huanwu Center for Fundamental Theory, Hefei, Anhui 230026, China}}
\date{\today}

\begin{document}

\begin{adjustwidth}{-40pt}{-40pt}
\maketitle
\end{adjustwidth}

\begin{abstract}
Fundamental principles of local quantum field theory or of quantum gravity can enforce consistency requirements on the space of consistent low-energy effective field theories. We survey the various techniques that have been used to put UV constraints on IR physics, including those from causality considerations in the form of S-matrix positivity and bootstrap bounds, scattering time delays, conformal field theory and holographic methods, together with those that arise from landscape/swampland criteria such as the weak gravity conjecture.
We review recent applications of these constraints to corrections to Standard Model physics, corrections to Einstein gravity, and cosmological theories and highlight promising future directions.
\end{abstract}

\snowmass

%\maketitle

\tableofcontents

\section{Introduction}

Low energy Effective Field Theories (EFTs) are the central tool in the theoretical description of low energy particle physics, cosmology and gravitational theories. Modern understanding has it that the Standard Model of particle physics and Einstein gravity should both be understood as leading terms in an EFT expansion. It is thus of paramount importance to understand the space of allowed low energy (IR) EFTs. The traditional perspective is that in constructing an EFT one should include every local operator consistent with the symmetries of the theory, suppressed by the appropriate power of the cutoff. This leads to a huge freedom in the space of allowed Wilson coefficients that EFT principles alone do not constrain. \\

Recently, there has been significant progress in understanding what is the space of low energy EFTs that admit a UV completion, either through string theory, quantum gravity or within field theory alone. Our goal in this review is to survey some of the most promising methods. One of the most startling results of recent work is the recognition that the space of allowed coefficients, far from being free, is often compact and significantly constrained in a manner anticipated by naturalness arguments. This follows in particular from a set of powerful S-matrix positivity and bootstrap bounds. Built into this is the assumption that the UV completion is
\begin{itemize}
\item Unitary
\item Local and Causal
\item Lorentz Invariant
\end{itemize}
Although string theory involves extended objects, it retains many of the analytic properties that encode the notion of locality in quantum field theory, and so remains amenable to this general framework. These same assumptions can be made in conformal field theories and can be used to place constraints on QFTs with interacting UV fixed points, and via holography, gravitational theories in Anti-de Sitter, and are incorporated in the powerful analytic conformal bootstrap methods (see the Snowmass white paper \cite{Hartman:2022zik}). \\

Separately from these formal considerations, explicit quantum gravity constructions provided by string theory provide guidance on the expected properties of IR EFTs. In recent years, these have been formalised as a number of conjectures believed to separate the landscape of consistent IR theories from the swampland of EFTs which do not admit a quantum gravity UV completion. Notable examples are the weak gravity conjecture, species bound, swampland distance conjecture and charge convexity conjecture. Putting such conjectures on the same footing as the more rigorous positivity and bootstrap bounds is a key target for future work.

\section{Constraints on IR Physics from UV QFT}\label{sec:QFT}

\subsection{Causality and Analyticity in QFT}
Causality is inherently a Lorentzian concept. In QFT, it requires commutators of local operators to vanish outside the light-cone:
\begin{equation}\label{label:causality}
[O_1(x), O_2(0)]=0\ , \qquad x^2>0\ ,
\end{equation}
where $x\in \mathbb{R}^{d-1, 1}$. Thus, causality is highly nontrivial in QFT, since it is an operator statement. For example, all correlators containing the commutator (\ref{label:causality}) must vanish exactly, leading to infinite sets of consistency conditions in QFT. These consistency conditions are encoded in the analytic structure of correlation functions.

A QFT can be equivalently defined by its Euclidean correlators. Lorentzian correlators of any ordering can then be obtained by analytically continuing these Euclidean correlators. Moreover, the analytic structure of Lorentzian correlators of unitary QFTs directly follows from the fact that Euclidean correlators are single-valued, permutation invariant functions of positions that do not have any branch cuts, provided all points remain Euclidean \cite{haag}. In particular, any Lorentzian correlator $\langle O_1(x_1)O_2(x_2)O_3(x_3) \cdots\rangle $ of local operators, as a  function of complexified $x_i$, is analytic in the domain \cite{haag,Hartman:2016lgu}:
\begin{equation}\label{con1}
\text{Im}\ x_1 \vartriangleleft \text{Im}\ x_2  \vartriangleleft \text{Im}\ x_3  \vartriangleleft \cdots\ ,
\end{equation}
where, $\mbox{Re}\ x_i \in \mathbb{R}^{d-1, 1}$ and $\mbox{Im}\ x_i \in \mathbb{R}^{d-1, 1}$. The symbol $x \vartriangleleft y$ represents that the point $y$ is in the future light-cone of point $x$. This analyticity condition is precisely a covariant version of the standard $i\epsilon$ prescription. Strictly speaking, the analyticity condition (\ref{con1}) is a necessary condition but not sufficient because Lorentzian correlators have an even larger (but more complicated) domain of analyticity, as described in \cite{haag}.

%\subsubsection{Applications in CFT and energy conditions}
The analyticity condition (\ref{con1}) is particularly useful in conformal field theory (CFT), where it has been rigorously established from convergence of the operator product expansion (OPE) \cite{Hartman:2015lfa} (see also \cite{Kravchuk:2020scc,Kravchuk:2021kwe}). In fact, in CFT, causality of four-point functions places nontrivial constraints on the spectrum of light operators, independent of the rest of the theory~\cite{Hartman:2015lfa,Hartman:2016dxc,Hofman:2016awc,Chowdhury:2017vel,Afkhami-Jeddi:2017idc,Chowdhury:2018uyv,Kundu:2020gkz}. Hence, these CFT constraints should be interpreted as IR constraints from UV consistency. Furthermore, the analyticity condition plays a crucial role in many of the important results in the conformal bootstrap; for a review, see~\cite{Poland:2018epd}.

More generally, the analyticity condition (\ref{con1}) led to the proof of the averaged null energy condition (ANEC) for all unitary, Lorentz-invariant QFTs with interacting conformal fixed points in the UV \cite{Hartman:2016lgu}.\footnote{The first general proof of the ANEC was found by using monotonicity of relative entropy \cite{Faulkner:2016mzt}. This suggests a curious connection between causality and quantum information in QFT. Both of these approaches were later combined to prove the quantum null energy condition in \cite{Balakrishnan:2017bjg}.} It is well-known that the ANEC is a nontrivial statement about QFT even in Minkowski spacetime without gravity. For example, the ANEC has been utilized in the conformal collider setup, obtaining rigorous bounds on CFT data \cite{Hofman:2008ar,Cordova:2017zej,Cordova:2017dhq,Meltzer:2018tnm,Jensen:2018rxu,Kologlu:2019mfz}.

However, for non-conformal QFTs it is technically challenging to compute position space correlators. So, an alternative, but not exactly equivalent, S-matrix based framework has been developed to study causality. We will discuss that next.

\subsection{Unitarity, Causality and Analyticity in S-Matrix}
% Andrew (Plan: Dispersion relations and unitarity. Brief summary: why does locality lead to analyticity? UV behavior (Froissart bound). Positivity from forward scattering~\cite{Adams:2006sv}. Extending away from the forward limit. Subtleties with massless particles.) \\

\begin{figure}[h]
\centering
\includegraphics[scale=0.5]{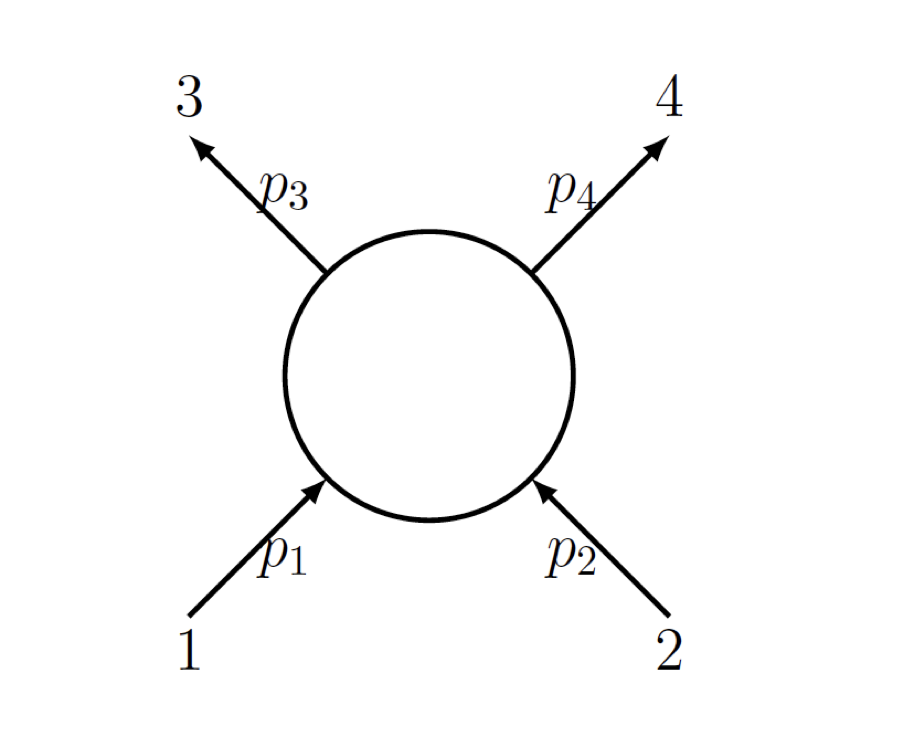}
\caption{ \label{fig:intro} \small 2-2 scattering amplitude}
\end{figure}
In this section we focus on UV-complete quantum field theory, i.e., on theories which approach a (possibly free) local conformal field theory in the ultraviolet for which gravity may be neglected. In the 1950s and 1960s, various rigorous theorems were proven demonstrating the connection between QFT causality, as encapsulated in the vanishing of commutators outside of the lightcone \eqref{label:causality}  and the analyticity of the S-matrix \cite{Symanzik1957DERIVATIONOD,lehmann1958analytic,Bremermann1958PROOFOD,bogoliubov1959introduction,Hepp_1964,bros1964some}.
 The inspiration for these proofs comes from the Kramers-Kronig relations which are dispersion relations for the refractive index. The requirement that the retarded propagator which is essentially determined by expectation values of $\theta(x^0-y^0) [O_1(x),O_2(y)]$ vanishes for $x^0<y^0$ can be translated into a statement about the analyticity of its Fourier transform via Titchmarsh's theorem \cite{Nussenzveig:1972tcd,Gell-Mann:1954ttj}, and hence analyticity of the refractive index $n(\omega)$. The Kramers-Kronig dispersion relations follow from Cauchy's theorem applied to the refractive index with a single subtraction.

In the case of relativistic QFT, the central object of consideration is the 2-2 scattering amplitude ${\cal A}(s,t) $ defined via the $T$-matrix $\langle p_3 , p_4 | \hat T | p_1, p_2 \rangle= {\cal A}(s,t) (2\pi)^4 \delta^4(p_1+p_2-p_3-p_4)$ and the first derivation of a dispersion relation from causality was in \cite{Gell-Mann:1954ttj}. Using the LSZ construction this scattering amplitude can be related to integrals of expectation values of retarded products $ \theta(x^0-y^0) \langle p_3 |  [\phi(x),\phi(y)]| p_1 \rangle$ \cite{Lehmann1957OnTF}. From the Jost-Lehmann-Dyson representation \cite{dyson1958integral,jost1957integral}, the scattering amplitude of identical mass particles can be shown to be analytic in $s$ up to poles at $s=m^2$ and $s=4m^2-t$, with a right hand branch cut beginning at $s=4m^2$, and a left hand branch cut beginning at $s=-t$ \cite{bogoliubov1959introduction,Hepp_1964}. These motivate the central tenets of the S-matrix program:
\begin{itemize}

\item Maximal Analyticity (Causality): The physical scattering amplitude is the boundary value of an analytic function of complexified Mandelstam variables modulo physical poles and cuts determined by on-shell unitarity.
\item Polynomial (Exponential) Boundedness (Locality): The growth of the scattering amplitude at large values in the complexified Mandelstam plane is bounded by an polynomial or a linear exponential. In the polynomial case this follows from the Wightman axioms.
\item Unitarity: $\hat S^{\dagger} \hat S = 1$.
\item Lorentz Invariance.
\item Crossing Symmetry: Physical region scattering amplitudes will be related under interchange of Mandelstam variables which can be achieved via a path in the complex Mandelstam plane.
\end{itemize}
Using Cauchy's theorem and assuming polynomial boundedness $|{\cal A}(s,t)|< |s|^N$ we have
\begin{equation}
     \frac{{\cal A}(s,t)}{s^N} = \frac{1}{2\pi i}\int_{{\cal C}} d s' \frac{{\cal A}(s',t)}{(s')^N(s'-s)} \,,
\end{equation}
for a contour ${\cal C}$ encircling the point $s'=s$ but not including $s'=0$ or any other poles/cuts. $N$ denotes the number of subtractions needed to account for the polynomial growth of the amplitude at infinity. Deforming the contour ${\cal C}$ to surround the physical cuts and poles and subtraction point $s'=0$, the following dispersion relation is obtained
\begin{equation}\label{dispersion}
    {\cal A}(s,t) = \frac{\lambda_s(t)}{m^2-s}+\frac{\lambda_u(t)}{m^2-u} + \sum_{n=0}^{N-1} a(t)s^n +  \frac{s^N}{\pi} \int_{4m^2}^{\infty} d \mu \frac{{\rm Abs}\,{\cal A}_s(\mu,t)}{\mu^{N} (\mu-s)} +  \frac{u^N}{\pi} \int_{4m^2}^{\infty} d \mu \frac{{\rm Abs}\,{\cal A}_u(\mu,t)}{\mu^{N} (\mu-u)}\,.
\end{equation}
The absorptive parts are given by the discontinuities of the right and left hand branch cuts\footnote{${\rm Abs}\,{\cal A}_s(s,t)\equiv \frac{1}{2i}\left( {\cal A}(s+i \epsilon,t)-{\cal A}(s-i \epsilon,t) \right)$ and ${\cal A}_u(s,t)={\cal A}_s(u,t)$.}. Martin \cite{Martin:1965jj} demonstrated, using unitarity, that the full validity of this dispersion relation could be extended to positive $t<4m^2$. Furthermore following Froissart \cite{Froissart:1961ux}, Jin and Martin \cite{PhysRev.135.B1375} proved that in theories with a mass gap, the scattering amplitude in this region is bounded
\begin{equation}
|{\cal A}(s,t)|< |s|^{\alpha(t)} \, ,  \quad  -|t_0| \le t< 4 m^2 \, ,
\end{equation}
with $\alpha(t)<2$,  which follows from unitarity and positivity of all $t$-derivatives of the absorptive parts. This fixes the number of subtractions to be $N=2$ both in the forward $t=0$ limit and non-forward limit $0 < t< 4 m^2$.

Recent developments in positivity bounds and S-matrix bootstrap methods begin with the dispersion relation \eqref{dispersion} often assumed applicable in the case of massless particles. Its validity in the case of massless particles is more subtle and relies on assuming there remains some Froissart-like or Regge bound on the growth of the amplitude over the whole complex plane, which while not rigorously proven is expected in string theory completions \cite{Chandorkar:2021viw} and quantum gravity \cite{Haring:2022cyf}. Recent work on positivity bounds was reinvigorated in \cite{Adams:2006sv} where it was shown that already in the forward limit $t=0$, the above dispersion relation implies positivity of all even $s$ derivatives of the pole subtracted amplitude
\begin{equation}\label{bounds1}
\partial_s^{2n}{\cal A}'(\bar s,0)  >0 \, , \quad n \ge 1 \, ,
\end{equation}
where $\bar s$ is a value on the real axis between the branch cuts.
The key recognition is that since these are constraints on the derivative expansion of the amplitude at low energies they imply constraints directly on the Wilson coefficients of operators in the low energy effective action. In particular for a scalar Goldstone EFT, the operator $c(\partial \phi)^4 $ or equivalently $ c \phi^2 \Box^2 \phi^2$ contributes to the scattering amplitude in the manner $c(s^2+t^2+u^2)$ and is forced to be positive $c>0$ by the requirement that $\partial_s^{2}{\cal{A}}'(\bar{s},0)  >0 $. The authors of \cite{Adams:2006sv} noted that this is consistent with naive causality considerations based on the speed of propagation. The higher order bounds \eqref{bounds1} similarly impose positivity of higher derivative operators of the form $\phi^2 \Box^k \phi^2$ for $k \ge 2$.

These linear bounds have been extended away from the forward limit using the positivity properties of the Legendre Polynomials $\partial_x^n P_{\ell}(1+x)|_{x=0}\ge 0$ \cite{deRham:2017avq,Pennington:1994kc,Vecchi:2007na,Manohar:2008tc,Nicolis:2009qm} or equivalently Gegenbauer polynomials $\partial_x^n C^{(D-3)/2}_{\ell}(1+x)|_{x=0}\ge 0$ in $D$ dimensions, giving an infinite number of bounds \cite{deRham:2017avq}. The bounds have further been extended to particles of arbitrary massive spins in \cite{Bellazzini:2016xrt,deRham:2017zjm}. In the massive spin case, care must be taken with the additional kinematic singularities and the non-trivial nature of crossing symmetry \cite{deRham:2017zjm,Hebbar:2020ukp}. A fruitful application of these bounds has been to effective theories of massive spin-2 states \cite{Cheung:2016yqr,Bonifacio:2016wcb,deRham:2017xox,Bellazzini:2017fep,deRham:2018qqo,Alberte:2019xfh,Alberte:2019zhd,Wang:2020xlt}. In particular, as noted in \cite{Cheung:2016yqr}, bounds from different polarizations impose a finite `island' of allowed Wilson coefficients. Similar bounds may be derived for lower and higher spins \cite{deRham:2017imi,Davighi:2021osh,Bellazzini:2019bzh}.

An important recent development was the recognition of the existence of nonlinear positivity bounds which incorporate the positivity properties of the Legendre/Gegenbauer polynomials in the `EFT-Hedron' \cite{Arkani-Hamed:2020blm,Chiang:2021ziz} together with the positivity of the moments of a positive distribution. For example, for any function which admits a dispersion relation with positive spectral representation
\begin{equation}
    A(s) = \int_{m_\mathrm{gap}^2}^{\infty} d \mu \frac{\rho(\mu)}{\mu-s} \, ,
\end{equation}
with $\rho(\mu) \ge 0$, defining $a_n = (n!)^{-1}\partial_s^n A(0)$ one can construct an $N \times N$ Hankel matrix $H_n$ whose components are $(H_n)_{pq}= a_{n+p+q}$ with $p,q=0,1, \dots, N-1$. Given any vector $V$ whose coefficients are $V_p = y^p$, i.e., $V=(1,y,y^2 \dots, y^{N-1})$, it is straightforward to show that
\begin{equation}
    V^T H_n V = \sum_{p,q=0}^{N-1} a_{n+p+q}y^p y^q =  \int_{m_\mathrm{gap}^2}^{\infty} d \mu \frac{\rho(\mu)}{\mu^{1+n}} \left( \sum_{p=0}^{N-1} \frac{y^p}{\mu^p}\right)^2 >0 \,,
\end{equation}
for all $y$, which implies positivity of the determinant of the Hankel matrix ${\rm Det}\left[ H_n \right] > 0$, for example $a_2a_0-a_1^2>0$.
This leads to an infinite number of nonlinear positivity bounds which incorporate the methodology of a Hausdorff/Stieltjes moment problem \cite{Bellazzini:2020cot} paralleling the 1970's positivity bounds of Common and Yndurain\cite{common1969properties,yndurain1969constraints,common1970some,common1970constraints}.  The EFT-Hedron \cite{Arkani-Hamed:2020blm} gives a geometric description of `positive geometry' of these bounds, closely related to that found in grassmannians and amplituhedra, giving powerful nonlinear constraints on Wilson coefficients at all orders.

The dispersion relation \eqref{dispersion}, while manifestly $s \leftrightarrow u$ crossing symmetric, is not manifestly $s \leftrightarrow  t$ crossing symmetric. Further imposing full crossing symmetry imposes a set of `null-constraints' which has tightened these nonlinear bounds \cite{Tolley:2020gtv,Caron-Huot:2020cmc,Du:2021byy}.
These constraints have been used to show how positivity bounds imply  upper and lower bounds on Wilson coefficients, or more generally compact `islands' of parameter space even for scalar theories, similar to the case of spin-2. In general, finding optimal positivity bounds can be formulated as a semi-definite program with a continuous decision variable, solvable by the SDPB package \cite{Simmons-Duffin:2015qma}, and the Wilson coefficients can be constrained to be ${\cal O}(1)$, putting the usual EFT naturalness assumption on a rigorous footing. In particular, \cite{Tolley:2020gtv} uses this to show that all softly broken Galileon theories are ruled out by standard positivity bounds. A useful technique introduced in \cite{Caron-Huot:2020cmc} demonstrates how null-constraints can be used to give special combinations of amplitudes (sum-rules) which are finite polynomials in Mandelstam invariants. The effect of IR loops on these type of bounds is discussed in \cite{Bellazzini:2021oaj}.

These two-sided positivity bounds or `island' bounds strongly overlap with bounds derived using the S-matrix bootstrap program which has a `primal' \cite{Paulos:2016fap,Paulos:2016but,Paulos:2017fhb,Homrich:2019cbt,Guerrieri:2020bto,Correia:2020xtr} and `dual' formulation \cite{Guerrieri:2020kcs,Cordova:2018uop,He:2021eqn,EliasMiro:2021nul,Guerrieri:2021tak} (see the S-matrix bootstrap Snowmass white paper \cite{Kruczenski:2022lot}).  The S-matrix bootstrap methods use the same dispersion relation \eqref{dispersion} as a starting point, but utilize the power of full unitarity for each partial wave (rather than positivity alone) together with a truncated crossing symmetric expansion for the scattering amplitude to determine numerical bounds on Wilson coefficients. In particular \cite{Guerrieri:2020bto} gives bounds on pion scattering closely related to those obtained via positivity methods \cite{Arkani-Hamed:2020blm,Bellazzini:2020cot,Tolley:2020gtv,Caron-Huot:2020cmc}. Entanglement/relative entropy has been incorporated into the bootstrap in \cite{bose2021selection,bose2020relative}. A related bootstrap approach which tries to fix the form of the amplitude from a knowledge of particle production is given in \cite{Tourkine:2021fqh}.

A reformulation of the standard fixed $t$ dispersion relation \eqref{dispersion} in a form in which crossing symmetry is manifest is given in \cite{Sinha:2020win}, which allows for an elegant alternative derivation of null-constraints and positivity bounds following from the Bieberbach conjecture \cite{Haldar:2021rri} with interesting relations to geometric function theory, and a proof of the two sided bounds on Wilson coefficients using the Markov brothers’ inequality \cite{Raman:2021pkf}.

\subsection{Unitarity and Causality Bounds for Gravitational EFTs}

A direct application of the previous positivity bounds to gravitational effective field theories, such as low energy string theories, is not possible due to the technical issues associated with a massless spin-2 state, the graviton. Firstly the masslessness of the graviton means there is no mass gap which plays a crucial role in several arguments, notably the proof of the Froissart-Jin-Martin bound and the analytic continuation of the partial wave expansion from $t<0$ to $t>0$. This is exacerbated in $D=4$ due to IR divergences. These problems can be ameliorated by working at tree-level in the low energy effective theory, i.e. not including light loops, which is sufficient for weakly coupled UV completions such as string theory.
Furthermore arguments can be given for `Froissart-like' Regge bounds for amplitudes for $t<0$ \cite{Chandorkar:2021viw,Haring:2022cyf}.

The second problem is that even at tree level the presence of a spin-2 massless pole undermines the assumption of a dispersion relation with two subtractions for $t \ge 0$ and yet the positivity properties of the Legendre/Gegenbauer polynomials only apply for $t \ge 0$. There is however no problem working with more than 2 subtractions, and a powerful set of bounds using the EFT-hedron and null-constraints can be derived for tree level EFT graviton scattering amplitudes \cite{Bern:2021ppb,Chowdhury:2021ynh, Chiang:2022jep}. The validity of the leading positivity bounds in the presence of a spin-2 pole was considered in \cite{Alberte:2020jsk,Alberte:2020bdz} where it was argued that the leading positivity bounds could be mildly violated in gravitational effective field theories by an $M_{\rm Pl}^{2}$ suppressed amount. This conjecture was confirmed by the beautiful results of \cite{Caron-Huot:2021rmr} where they used the power of the null-constraints together with the scattering amplitude in impact parameter space, or equivalently at $-\Lambda^2<t<0$ to define a set of positive integrals which could be used to constrain the leading $s^2$ terms in the scattering amplitudes even in the presence of the massless spin-2 pole. These bounds have been put on the leading order corrections to Einstein gravity in \cite{Caron-Huot:2022ugt}, complementing the results of \cite{Bern:2021ppb, Chowdhury:2021ynh, Chiang:2022jep}. An interesting observation is that strong bounds can be obtained by assuming that the lower spin partial waves dominate in the dispersion relation, as shown in explicit examples, via  geometric function theory or via SDP numerically \cite{Bern:2021ppb, Chowdhury:2021ynh, Caron-Huot:2022ugt, Chiang:2022jep}.

A distinct way to deal with the $t$-channel spin-2 poles is to subtract out the high energy Regge behaviour which it implies in the dispersion relation to give a new dispersion relation with only two subtractions \cite{Hamada:2018dde,Tokuda:2020mlf,Herrero-Valea:2020wxz}. These arguments confirm the expectation that the leading positivity bounds are mildly violated in the presence of the graviton by an amount determined by the Regge behaviour. In \cite{Alberte:2021dnj} it is shown that positivity bounds for specific indefinite polarization photon-graviton scattering amplitudes could be used to impose constraints on the Regge behaviour which are strongly dependent on the IR physics, reversing the usual logic.

Following the numerical S-matrix bootstrap program, bounds on the gravitational EFTs have been considered in \cite{Guerrieri:2021ivu} where it is shown that the constraints on the allowed Wilson coefficient of the leading non-trivial corrections to $D=10$ maximal supergravity fit well with the range of allowed values expected in string theory  (see the Snowmass white paper on bootstrapping string theory \cite{Gopakumar:2022kof}).

\subsection{Causality Constraints from Propagation Speed and Time-delay}

\paragraph{Low-energy Phase-velocity (Super)luminality:}
It is typically argued that when propagating on media or on backgrounds that spontaneously break Lorentz-invariance, the low-energy phase velocity can be superluminal, so long as the group velocity remains (sub)luminal. In reality, neither statements are technically correct and examples in nature have manifested instances of superluminal group and phase velocity while remaining in perfect agreement with causality \cite{PhysRevLett.93.203902}.
In practice, causality demands that the speed of propagation of the information, i.e., the front velocity, be luminal (where the notion of luminality is indicated by the geometry seen by those high-frequency modes). Since the front velocity is the infinite frequency limit of the phase velocity, causality is intrinsically a property of the UV which manifest itself as constraints on the IR. Embedded in the Kramers-Kronig relations, the statement of unitarity and that of analyticity dictate that the phase velocity cannot decrease with frequency in standard low-energy EFTs (as is the case for light) which explains why a superluminal low-frequency phase or group velocity is typically associated with a violation of causality (unless one allows for ``exhibiting gain" where for instance the laser is allowed to take energy from the system and the condition of unitarity is effectively violated in the reduced phase space \cite{PhysRevLett.93.203902}). Focusing on relativistic IR field theories that respect unitarity, and analyticity and which admit a dispersion relation (or Kramers-Kronig relations) with two subtractions, there is therefore a direct link between causality violation and a superluminal low-energy phase velocity.

The connection between low-energy superluminality and a violation of either unitarity or causality in the UV is manifest for instance in the Goldstone scalar model $\phi$ in flat spacetime made clear in \cite{Adams:2006sv}. A low-energy EFT operator of the form $c(\partial \phi)^4$ can be thought as deriving from the integration of a heavy mode $H$ in the partial UV with interaction $\alpha H (\partial \phi)^2$, with $c\sim \alpha^2$. Embedding the low-energy EFT in a unitary (partial) UV completion, therefore demands $\alpha$ to be real and hence positivity of the coefficient $c>0$, which is precisely what is inferred from the previous positivity constraints irrespectively of the precise details of the UV completion. From a low-energy EFT point of view a similar conclusion could have been drawn by simply inspecting the low-frequency phase (or group) velocity of the Goldstone mode on a background that spontaneously breaks Lorentz invariance. Irrespectively of the details of the background, subluminality demands $c>0$, in agreement with the commonly accepted connection between superluminality and the violation of either causality or unitarity \cite{Adams:2006sv}. However in more generic EFTs, particularly gravitational ones, the previous arguments are more subtle. In the gravitational context, the notion of speed is frame-dependent and observables related to scattering (phase shift or time delay) are thus better probes of how causality manifests itself in a low-energy EFT \cite{Eisenbud:1948paa,Wigner:1955zz,Smith:1960zza,Martin:1976iw,de2002time}.

\paragraph{Time Delay: Asymptotic and Infrared Causality}
In generic low-energy EFTs, including gravitational ones, causality can be imposed by demanding that upon scattering a wave of asymptotic energy $\omega$ in an asymptotically flat spacetime, the net (Eisenbud-Wigner) scattering time delay $\Delta T^{\rm net}$ should be bounded by $ \Delta T^{\rm net} \gtrsim -\omega^{-1}$. This criterion is often referred to as asymptotic (sub)luminality or asymptotic causality \cite{Gao:2000ga,Camanho:2014apa} and becomes particular powerful when considering scattering around shock wave or plane-wave spacetimes \cite{Camanho:2014apa}. This condition refers to the absence of superluminality as compared to the asymptotic flat Minkowski metric and is consistent with expectations from a Lorentz invariant UV completion. According to the criterion of asymptotic causality, if the net time delay is negative,  the scattered wave enjoys a net time advance $\Delta T^{\rm net, adv}= -\Delta T^{\rm net}$ as compared to the asymptotic Minksowki spacetime. When this time advance is resolvable it would be a clear violation of causality, which can be used to put constraints on Wilson coefficients in the EFT. Applications have been given in \cite{Camanho:2016opx,Goon:2016une,Hinterbichler:2017qcl,Hinterbichler:2017qyt,AccettulliHuber:2020oou,Bellazzini:2021shn}.

While asymptotic causality is likely a necessary criterion for a theory to admit a Lorentz invariant UV completion, it is not clear if it is manifests the full implications of causality. Fundamental definitions of causality such as \eqref{label:causality} are defined locally at least for non-gravitational theories, and require no mention of the asymptotic geometry. For gravitational theories, the equivalence principle dictates that locally, the high-frequency modes which determine the support of retarded propagators can only be sensitive to the local inertial frame and not to the asymptotic one. A more refined criterion for causality, ``infrared causality'', which incorporates this is given in \cite{Hollowood:2015elj,deRham:2020zyh,Chen:2021bvg,deRham:2021bll}, which dictates that the net time delay $\Delta T^{\rm net}$ of low-energy EFT modes should be bounded by that inferred from the background geometry, or Shapiro time delay $\Delta T^{g}$ as seen by the high-frequency modes, in the sense that $\Delta T^{\rm net}>\Delta T^g-\mathcal{O}(1)\omega^{-1}$. This definition is consistent \cite{Chen:2021bvg,deRham:2021bll} with expectations of gravitational positivity bounds \cite{Caron-Huot:2022ugt} and with the QFT in a fixed background decoupling limit.

The statements of asymptotic and infrared causality have already proven to provide insightful bounds on IR gravitational theories from the requirement of a causal UV completion. Within higher-dimensional gravity, the effect of the Gauss-Bonnet terms were constrained using these approaches~\cite{Camanho:2014apa,Chen:2021bvg}, while constraints on generic EFTs of gravity were considered in \cite{AccettulliHuber:2020oou,deRham:2021bll}, and QED in curved spacetime in~\cite{Hollowood:2010xh,Hollowood:2012as,Hollowood:2015elj,Goon:2016une,deRham:2020zyh,Bellazzini:2021shn}, and applied to constrain other classes of low-energy EFTs \cite{Hinterbichler:2017qcl,Hinterbichler:2017qyt}.
In \cite{Gruzinov:2006ie}, it was proven that the requirement of subluminality in the EFT of gravity required the coefficient of the parity-invariant dimension-8 operators (Riem$^{4}$) to be positive, which was later proven to be fully consistent with positivity bounds requirements as imposed by the UV \cite{Bellazzini:2015cra}. In  \cite{deRham:2021bll}, the requirement of infrared causality was further shown to demand additional conditions on these dimension-8 operators consistent with what was later found using positivity bounds at finite impact parameter \cite{Caron-Huot:2022ugt}. Moreover, when applied to scattering about black holes,  the statement of infrared causality was also shown to impose that the sign of the dimension-6 operators (Riem$^3$) of the EFT of gravity must be nonnegative, a statement that is in agreement with all known realizations of string theory and yet requires no knowledge of string theory to be inferred and should be generic to any unitary and causal tree-level weakly-coupled UV completion.

While positivity bounds can be derived more rigorously and typically allow to extract a greater level of information from the UV to the IR, their application fails in some situations, either due to the presence of a gravitational exchange or when considering situations where the spontaneous breaking of Lorentz invariance is intrinsically related to the EFT considered, as is the case for some EFTs of cosmology (see the Snowmass white paper on EFT for cosmology~\cite{whitepapereftcosmology}). In such situations, a more local requirement such as infrared causality which does not rely on the assumption of asymptotically Minkowski spacetime can be used to make progress and constrain these classes of EFTs as was for instance demonstrated in \cite{deRham:2020zyh}. The relation between causality and positivity in higher-spin theories can also be considered and further used to constrain those EFTs in the IR; see~\cite{Porrati:2008ha,Caron-Huot:2016icg,Hinterbichler:2017qcl,Afkhami-Jeddi:2018apj,Kaplan:2019soo, Bellazzini:2019bzh,Kaplan:2020ldi,Kaplan:2020tdz}.

\subsection{Applications to SMEFT}

In the absence of new particle discoveries yet at the LHC, the Standard Model Effective Field Theory (SMEFT) \cite{Weinberg:1978kz, Buchmuller:1985jz, Leung:1984ni} (see the Snowmass white paper on the SMEFT) has been becoming an increasingly popular tool to systematically parametrize new physics beyond the Standard Model. A challenge in the SMEFT approach is that, due to the large number of degrees of freedom already in the Standard Model, there are many effective operators in the SMEFT \cite{Henning:2015alf, Lehman:2015coa}, so its naive parameter space, even truncated at dimension-8, is extremely large \cite{Grzadkowski:2010es, Li:2020gnx, Murphy:2020rsh}, which hinders experimental searches and theoretical interpretation of the data. Therefore, it is very useful to constrain the SMEFT parameter space by reliable theoretical priors such as positivity bounds, and indeed, positivity bounds do eliminate large chunks of the naive parameter space \cite{Zhang:2018shp, Bi:2019phv, Bellazzini:2018paj, Remmen:2019cyz, Zhang:2020jyn, Yamashita:2020gtt, Fuks:2020ujk, Remmen:2020vts, Li:2021lpe}. For example, in vector boson scattering that probes anomalous quartic gauge boson couplings, only about 2\% of the naive parameter space of the 18 relevant electroweak dimension-8 operators satisfies elastic positivity bounds (the quadratic terms of the dimension-6 coefficients also enter the positivity bounds, but they always contribute negatively to the bounds, so neglecting them gives conservative bounds) \cite{Zhang:2018shp, Bi:2019phv} (see also \cite{Vecchi:2007na, Remmen:2019cyz}); Generalized elastic bounds that mix different particle species lead to even stronger constraints --- only about 0.7\% being consistent with positivity already for the transversal gauge bosons \cite{Yamashita:2020gtt}. For the fermionic sector of the SMEFT, generalized elastic bounds imply that any flavor-violating Wilson coefficients are interestingly bounded above by the flavor-conserving ones, which, combined with the LEP data, can already preclude certain operators targeted by the upcoming Mu3e experiments \cite{Remmen:2020vts}. Assuming a minimal flavor violation structure, this direction is further explored by \cite{Bonnefoy:2020yee}. Positivity bounds on the SMEFT four-Higgs operators have been evaluated to one loop level \cite{Chala:2021wpj}, in which case it would be more appropriate to formulate the positivity bounds in terms of physical observables involving the amplitudes rather than directly in terms of the Wilson coefficients. Reversing the argument, positivity bounds on the SMEFT can also be used to test the fundamental principles of the S-matrix theory in future electron-positron colliders \cite{Fuks:2020ujk}, with a particularly clear channel identified in \cite{Gu:2020ldn}, (see \cite{Distler:2006if} for an earlier work in the context of electroweak chiral Lagrangian). Possible uses of dispersion relations/sum rules to constrain dimension-6 operators, with extra assumptions, have also been discussed in \cite{Adams:2008hp, Remmen:2020uze, Gu:2020thj, Davighi:2021osh, Azatov:2021ygj}. Various positivity bounds have also been applied to constrain chiral perturbation theory \cite{Pham:1985cr, Ananthanarayan:1994hf, Pennington:1994kc, Dita:1998mh, Manohar:2008tc, Mateu:2008gv, Sanz-Cillero:2013ipa, Du:2016tgp, Bellazzini:2017bkb, Wang:2020jxr, Zahed:2021fkp, Alvarez:2021kpq} and the Euler–Heisenberg EFT \cite{Henriksson:2021ymi, Chowdhury:2021ynh}.

In the context of the SMEFT, the most relevant positivity bounds are the leading forward limit bounds, which already come in at the order of dimension-8. For identical particle scattering, this means that the coefficient in front of $s^2$ is positive \cite{Adams:2006sv}; In the presence of multiple fields such as in the SMEFT, generalized elastic bounds are essentially the same after superimposing different fields, which still uses the standard optical theorem. However, one can extract more information by utilizing the generalized optical theorem in the dispersion relation, in which case it is more convenient to view the optimal positivity bounds as forming a convex cone \cite{Zhang:2020jyn}. %(It is worth noting that positivity from generalized optical theorem is essential to prove some versions of the weak gravity conjecture on black hole backgrounds \cite{Arkani-Hamed:2021ajd}.)
Additionally, the dual cone of this positivity cone is a spectrahedron (an intersection between the positive semi-definite matrix cone and an affine linear space), which means that the optimal positivity bounds can be efficiently obtained by semi-definite programming \cite{Li:2021lpe}, which is crucial for the SMEFT as it has a large number of fields. An interesting feature of the positivity cone is that its extremal rays are irreps of the SMEFT symmetries, which provides a group theoretical method to find the optimal bounds. In tree-level UV completions, an extremal ray corresponds to a UV particle with fixed SMEFT quantum numbers, while a ray inside the cone, on the other hand, would indicate presence of multiple UV states. Therefore, positivity bounds can provide important information to reverse-engineer the UV model from the SMEFT data \cite{Zhang:2020jyn, Fuks:2020ujk, Bellazzini:2014waa, Trott:2020ebl, Zhang:2021eeo}. On the other hand, confronting the experimental data with positivity bounds on dimension-8 operators can also be used to put exclusion limits on relevant BSM physics unambiguously, while excluding dimension-6 operators to a certain energy scale would be ambiguous. This is because a dimension-6 coefficient can vanish due to cancellations between multiple UV states, but that is not possible for a dimension-8 coefficient \cite{Fuks:2020ujk, Zhang:2021eeo}. An intriguing observation is that among the seesaw mechanisms only the heavy right-handed neutrino in the type-I model reaches the extremity of the positivity cone \cite{Li:2022tcz}, which means that the SMEFT can be used to pin down the seesaw mechanism if it is indeed type-I. All of these provide an important motivation to go beyond dimension-6 operators in the SMEFT, and measure and analyze the dimension-8 operators, a direction that is gaining pace recently \cite{Ellis:2017edi,  Bellazzini:2017bkb, Ellis:2018cos, Hays:2018zze, Bellazzini:2018paj, Ellis:2019zex, Ellis:2020ljj, CMS:2020ioi, CMS:2020meo, Gu:2020ldn, Fuks:2020ujk, Corbett:2021eux, Alioli:2020kez}.

\subsection{Applications to the EFT of Cosmology}

The direct application of S-matrix positivity bound/bootstrap methods to cosmology is problematic due to the lack of Lorentz invariance/crossing symmetry and the possible absence of an S-matrix. Since these symmetries are all broken at low energies and the positivity bounds rely on the properties at high energies, it may be sufficient to realise positivity in an approximate sense as in the application to EFTs of inflation in \cite{Baumann:2015nta} (see the Snowmass white paper on EFT for cosmology~\cite{whitepapereftcosmology}). One promising approach is to focus on the approximate S-matrix that from the broken boosts \cite{Grall:2021xxm} where dispersion relations may nevertheless be derived. An alternative approach may be to rely on the methodology of the cosmological bootstrap \cite{Arkani-Hamed:2018kmz,Baumann:2019oyu,Baumann:2020dch} in the hope that full power of the symmetries and crossing in de Sitter may be used.

A more pragmatic approach to constrain cosmological EFTs is to impose the constraints on Wilson coefficients defined around the Minkowski vacuum and translate this into constraints on the Wilson coefficients around cosmological backgrounds \cite{Melville:2019wyy,deRham:2021fpu,Traykova:2021hbr,Kim:2019wjo,Herrero-Valea:2019hde,Ye:2019oxx}.

\section{Constraints on Gravitational EFTs in AdS from CFT}\label{section:CFT}

%Sandipan

%+ connection with S-matrix bootstrap

%How causality, other properties of boundary QFT translate into constraints on the bulk EFT of a QG theory in AdS. Interface with bootstrap program.

UV consistency also imposes strict constraints on EFTs in anti-de Sitter (AdS) spacetime. Physically, these AdS bounds are very similar to the flat space bounds of section \ref{sec:QFT}, since local high energy scattering processes are expected to be insensitive to the spacetime curvature. However, EFTs in AdS enjoy  certain conceptual advantages mainly because of the dual CFT description. This allows us to derive AdS bounds directly from CFT axioms.

Originally, it was conjectured in~\cite{Heemskerk:2009pn} that any EFT in AdS$_D$ (with a large radius) has a dual CFT description in $D-1$ spacetime dimensions, where the dual CFT has the following properties:
\begin{itemize}
\item{The central charge\footnote{More precisely, $c_T$ is the coefficient of the CFT stress tensor two-point function. For gauge theories, the large $c_T$ limit is equivalent to the large-$N$ limit.} $c_T$ is large: $c_T\gg 1$,}
\item{A sparse spectrum: the lightest single trace operator with spin $\ell > 2$ has dimension
\begin{equation}
\Delta_\mathrm{gap}\gg 1\ .
\end{equation}
}
\end{itemize}
In fact, a scalar version of the conjecture was also proved in \cite{Heemskerk:2009pn} by  using the conformal bootstrap. Later, this idea was further studied and developed in \cite{Heemskerk:2010ty,Fitzpatrick:2010zm,Penedones:2010ue,ElShowk:2011ag,Fitzpatrick:2011ia,Fitzpatrick:2011hu,Fitzpatrick:2011dm,Fitzpatrick:2012cg,Goncalves:2014rfa,Alday:2014tsa,Hijano:2015zsa,Aharony:2016dwx}, implying that CFTs in this class, irrespective of their microscopic details, admit a dual holographic description in terms of EFTs in AdS.

So, now one can ask the dual question: what EFTs in AdS$_D$ cannot be embedded into a UV theory that is dual to a CFT$_{D-1}$ obeying the usual CFT axioms? This question has been explored extensively in recent years by leveraging the huge advancement in constraining the space of consistent CFTs from the conformal bootstrap \cite{Hofman:2008ar,Hartman:2015lfa,Afkhami-Jeddi:2016ntf,Paulos:2016fap,Paulos:2016but,Kulaxizi:2017ixa,Costa:2017twz,Afkhami-Jeddi:2017rmx,Paulos:2017fhb,Cordova:2017zej,Meltzer:2017rtf,Afkhami-Jeddi:2018own,Afkhami-Jeddi:2018apj,Homrich:2019cbt,Kaplan:2019soo,Haldar:2019prg,Conlon:2020wmc,Komatsu:2020sag,Kundu:2020bdn,Hijano:2020szl,Alday:2021odx,Kundu:2021qpi,Caron-Huot:2021enk}.

\subsection{Constraining 3-pt Interactions}
From this point of view, the constraints on gravitational three-point interactions of~\cite{Camanho:2014apa} can be derived directly  by constraining the space of consistent CFTs. In particular, the causality condition (\ref{con1}) for the dual CFT with large central charge and a sparse spectrum implies that the CFT three-point function $\langle TTT\rangle$ of the stress tensor must have a very specific structure ($D>4$) \cite{Afkhami-Jeddi:2016ntf,Afkhami-Jeddi:2017rmx,Afkhami-Jeddi:2018own}:\footnote{In AdS$_4$, the subleading terms are suppressed by $\frac{\ln \Delta_\mathrm{gap}}{\Delta_\mathrm{gap}^4}$ \cite{Afkhami-Jeddi:2018own}.}
\begin{align}\label{eq:einstein}
\langle T_{\mu_1 \nu_1}T_{\mu_2\nu_2}T_{\mu_3\nu_3}\rangle_\mathrm{CFT} = \langle h_{\mu_1 \nu_1}h_{\mu_2\nu_2}h_{\mu_3\nu_3}\rangle_\mathrm{Einstein}+\mathcal{O}\left(\frac{1}{\Delta_\mathrm{gap}^2}\right)\ .
\end{align}
The first term on the right-hand side is the CFT$_{D-1}$ correlator computed from the graviton 3-pt function of the Einstein action in the  bulk AdS$_D$ and the subleading terms are suppressed by $\Delta_\mathrm{gap}$. This constraint is precisely the CFT version of the constraint of  \cite{Camanho:2014apa}.  The same constraint was rederived and extended by imposing unitarity in the dual CFT in \cite{Kulaxizi:2017ixa,Costa:2017twz} and from the vanishing of the commutator of ANEC operators in \cite{Belin:2019mnx,Kologlu:2019bco}.

The above argument was later extended to derive rigorous bounds on 3-point functions $\langle T O_1 O_2\rangle$ of the stress tensor with various operators for CFTs with large central charge
and a sparse spectrum. In the AdS$_D$ language, these CFT constraints translate into the statement that all higher derivative interactions in the low energy effective action must be suppressed
by the new physics scale. In particular,  it was shown that the CFT dual of a bulk derivative is $1/\Delta_\mathrm{gap}$ in $D\ge 5$ \cite{Meltzer:2017rtf,Afkhami-Jeddi:2018own}. However, there is a logarithmic violation of this simple relationship between the bulk derivative and $\Delta_\mathrm{gap}$ in AdS$_4$ \cite{Afkhami-Jeddi:2018own}.

More recently, it has been shown that CFTs with large central charge and a sparse spectrum exhibit many other universal features, irrespective of their microscopic details (for example see \cite{Alday:2017gde,Kulaxizi:2018dxo,Kulaxizi:2019tkd,Fitzpatrick:2019efk,Huang:2021hye,Meltzer:2021bmb,Li:2021jfh}). However, a complete understanding of this class of CFTs  directly from the conformal bootstrap axioms is still an open question.

\subsection{Constraining 4-pt Interactions}
The conceptual advantage of the CFT-based argument becomes obvious when we study 4-pt interactions in EFT. To see that let us again consider the bound on the $\phi^2\Box^2\phi^2$ interaction of \cite{Adams:2006sv} (see section \ref{sec:QFT}). This bound (and many other similar bounds) was derived under the assumption that the $2\rightarrow 2$ scattering amplitude obeys (i) analyticity (in the usual regime), (ii) partial wave unitarity, (iii) crossing symmetry, and (iv) Regge boundedness conditions, even in the UV. These assumptions are well-motivated, however, they have not yet been rigorously established, especially for massless particles.

Another unsatisfying aspect of the original EFT argument is that the inclusion of dynamical gravity is technically challenging because of the graviton pole in the $2\rightarrow 2$ scattering amplitude. Moreover, it is unclear whether the Regge boundedness condition is valid in the presence of dynamical gravity. Although, it has recently been shown that some of these loopholes can  be bypassed by studying scattering amplitudes at finite impact parameter \cite{Caron-Huot:2021rmr, Caron-Huot:2022ugt}.

On the other hand, the sign constraint on the $\phi^2\Box^2 \phi^2$ coupling  can be  derived in AdS  in a conceptually cleaner way by using techniques from the conformal bootstrap. This EFT in AdS$_D$ is dual to an interacting CFT$_{D-1}$ which is obtained by deforming operator dimensions and OPE coefficients of a generalized free theory. When gravity is non-dynamical, the stress tensor of the dual CFT must decouple from the low energy spectrum. In other words, we are in the limit of large central charge $c_T\rightarrow \infty$ with $\Delta_\mathrm{gap}$ fixed (but large).
In this CFT setup, the sign constraint on the $\phi^2\Box^2 \phi^2$ coupling was derived in \cite{Hartman:2015lfa} directly from analyticity, positivity, and crossing symmetry of CFT 4-point correlators. These properties are well-established, since they follow from the conformal bootstrap axioms.
More recently, the CFT argument has been extended to impose precise bounds\footnote{It should also be noted that these CFT constraints are closely related to the chaos bounds of \cite{Maldacena:2015waa, Kundu:2021qcx, Kundu:2021mex}.} on the coupling constants of higher derivative interactions $\phi^2\Box^k\phi^2$ for all even $k\ge 2$ \cite{Kundu:2021qpi} (see also \cite{Caron-Huot:2021enk}). Another clear advantage of the CFT-based argument is that the inclusion of dynamical gravity is a rather trivial generalization. Now in the dual CFT, the central charge $c_T \gg \Delta_{\mathrm{gap}}\gg 1$ is large but finite. Using this CFT setup it was shown in \cite{Kundu:2021qpi} that dynamical gravity only affects constraints involving the $\phi^2\Box^2\phi^2$ interaction which now can have a negative coupling constant. This is also consistent with the flat space EFT results of \cite{Alberte:2020jsk,Alberte:2020bdz,Caron-Huot:2021rmr}.

So, the space of consistent EFTs in AdS can be analyzed in a conceptually clean way by utilizing the dual CFT description. It is only natural to expect that this CFT-based approach, in general, will play an important role in the Swampland program.

\section{Constraints on IR Physics from Quantum Gravity}

\subsection{The Landscape and the Swampland}

The space of effective quantum field theories is vast, and we can readily enumerate large classes of them. For example, we can postulate the existence any gauge group we like, and add matter fields transforming in various representations, subject to anomaly cancelation conditions. The space of consistent theories of quantum gravity appears to be much more restricted. Despite the apparent existence of a vast Landscape of string theory vacua, for instance, it is difficult or impossible to find vacua with arbitrarily large gauge groups, or with light matter fields in large representations of the gauge group. Conversely, string theory vacua often predict the existence of particles that need not exist in QFT, such as moduli and axion fields, hidden sector gauge groups, or Stueckelberg $U(1)$ extensions of the Standard Model gauge group (see, e.g.,~\cite{Halverson:2018vbo} and the Snowmass white paper~\cite{whitepaperstringconstructions}) In restricted settings, like intersecting D-brane models in Type II string theory~\cite{Blumenhagen:2005mu}, it is relatively straightforward to understand some consistency constraints on low-energy particle physics. For instance, matter fields transform in two-index tensor representations of gauge groups because an open string has two endpoints, each of which must be attached to a stack of D-branes. Tadpole cancelation imposes constraints that are stronger than, but closely related to, anomaly cancelation in QFT. For example, the ability to nucleate a brane-antibrane pair and extend the gauge group implies that the matter  content of the theory is constrained by anomaly cancelation within an extended gauge group~\cite{Halverson:2013ska}. Thus, within specific classes of quantum gravity theories, UV constraints on IR field theory are rather well understood. The situation becomes more complicated in the setting of F-theory (the nonperturbative limit of the Type IIB string theory), where the space of theories is still being understood. Thus, studies of {\em specific} theories of quantum gravity leave us with the question: to what  extent are the  common features that are found an artifact of looking under particular (stringy, often highly supersymmetric or higher-dimensional) lampposts? Are there first-principles arguments, like those we have discussed in quantum field theory, for how universal properties of the UV completion constrain the infrared effective theory?

One perspective on this problem is that, in parallel to the difficult work of gradually enlarging the Landscape of known, consistent theories of quantum gravity, we can identify regions of  the complementary ``Swampland,'' the space of effective quantum field theories that can not be embedded in consistent theories of quantum gravity~\cite{Vafa:2005ui}. (See~\cite{Brennan:2017rbf, Palti:2019pca, vanBeest:2021lhn, Grana:2021zvf, Harlow:2022gzl} for reviews.) Since we do not have a rigorous characterization of what quantum gravity {\em is}, in general, many of these ideas necessarily have a conjectural character. On the other hand, a great deal of evidence stands behind some of them, especially some of the oldest ideas, which predate the Swampland terminology.

One of the most significant claims, which underpins a large portion of all work in the Swampland program, is that theories of quantum gravity do not have global symmetries. This stands in stark contrast to quantum field theory, where global symmetries are a powerful organizing principle. However, it is consistent with the effective field theory perspective that good approximate  global  symmetries can arise as infrared accidents, as with baryon number in the real world, which can be broken by dimension-six operators in the Standard Model. In a gravitational theory, it is expected that such accidental symmetries can  be violated by black holes or wormholes~\cite{Hawking:1974sw, Zeldovich:1976vq, Giddings:1988cx, Abbott:1989jw, Banks:2010zn}. It is now generally expected that  quantum gravity theories violate all global symmetries, including generalized ($p$-form)  global symmetries~\cite{Gaiotto:2014kfa} or even more general notions of symmetry~\cite{McNamara:2019rup, Rudelius:2020orz, Heidenreich:2021tna}.

The mere statement that a symmetry is broken is not very useful on its own. It is important to ask how badly the symmetry is broken in the IR EFT. We expect that consistent theories of quantum gravity can allow symmetry violation  that is exponentially small in some parameter. For example, an axion field may have an approximate shift symmetry broken only by instanton effects of order  $\exp(-8\pi^2/g^2)$, where $g  \ll 1$  is a gauge coupling. However, making $g$ extremely small should come at a cost. The emerging picture, from studies of string theory examples, abstract  black hole arguments, and EFT arguments based on the idea  that weak coupling emerges from a strongly-coupled UV scale~\cite{Harlow:2015lma, Heidenreich:2017sim, Grimm:2018ohb} where all symmetries are badly broken~\cite{Cordova:2022rer}, is that any quantum gravity theory with an approximate global symmetry also has a UV cutoff scale that is low relative to the Planck scale: $\Lambda_\mathrm{QG} \ll M_\mathrm{Pl}$. At the scale $\Lambda_\mathrm{QG}$, local QFT is no longer a good approximation to the dynamics. In particular, we expect that any  semiclassical black hole in the theory will have radius $R_\mathrm{BH} \gtrsim \Lambda_\mathrm{QG}^{-1}$. The generic expectation that black holes are a source of global symmetry violation, then, suggests that the minimum size of symmetry-violating effects should be of order $\exp(-S_\mathrm{BH;min}) \sim \exp(- 8\pi^2 M_\mathrm{Pl}^2/\Lambda_\mathrm{QG}^2)$~\cite{Fichet:2019ugl, Daus:2020vtf}. In the aforementioned case of axions interacting via instantons, for example, this implies that $\Lambda_\mathrm{QG} \lesssim g M_\mathrm{Pl}$, which can also be argued from a different viewpoint~\cite{Heidenreich:2021yda}, related to the Weak Gravity Conjecture.

The Weak Gravity Conjecture (WGC) is a precise statement of how quantum gravity responds to the approximate global symmetry arising in the weak coupling ($g \to 0$) limit  of a gauge theory~\cite{ArkaniHamed:2006dz}. It argues that there must exist {\em superextremal} particles charged under a gauge field, i.e., particles with $m \leq \gamma q g M_\mathrm{Pl}$ where $q$ is the charge, $g$ is the gauge coupling, and $\gamma$ is an $O(1)$ constant determined by the extremality bound satisfied by black holes. For such particles, gravity is a weaker force than electromagnetic repulsion. By applying this bound to magnetic monopoles (with $g \to 2\pi/g$ by Dirac quantization) and appealing to the fact that new physics must enter at the length scale $(g^2 m_\mathrm{mon})^{-1}$ corresponding to the classical monopole radius, it was argued that the WGC implies a UV cutoff $\Lambda \lesssim g M_\mathrm{Pl}$. The last several years have seen intense study of the WGC and how it is obeyed in many corners of the string theory landscape. All known examples in fact obey much stronger bounds, the Tower WGC or Sublattice WGC, which postulate not just a single charged particle but an infinite tower of particles of different charge $q$, every one of which obeys the bound $m \leq \gamma q g M_\mathrm{Pl}$~\cite{Heidenreich:2015nta, Heidenreich:2016aqi, Montero:2016tif, Andriolo:2018lvp}.

The Tower WGC has a close affinity with the older Swampland Distance Conjecture  (SDC)~\cite{Ooguri:2006in}, which argues that whenever a scalar field $\phi$ in a quantum gravity theory traverses a large geodesic distance in field space, an infinite tower of modes becomes light, with masses $m \propto \exp(- c \Delta \phi / M_\mathrm{Pl})$. Here $\Delta \phi$ is the field-space distance and $c$ is an $O(1)$ constant~\cite{Klaewer:2016kiy}. In string theory, gauge couplings are controlled by expectation values of moduli fields, so the $g \to 0$ limit is also an infinite-distance limit, and the tower of charged particles predicted by the Tower WGC also obeys the SDC in examples. The existence of a tower of states also implies that the UV cutoff becomes low as $\Delta \phi$ increases. This happens in accordance with an even older proposal, the Species Bound, which asserts that any quantum gravity theory in $D$ spacetime dimensions with $N$ weakly coupled fields below the cutoff scale has a low UV cutoff $\Lambda_\mathrm{QG} \lesssim N^{-\frac{1}{D-2}} M_\mathrm{Pl}$~\cite{Veneziano:2001ah, Dvali:2007hz}. This is because the particles run in loops and contribute to the growth of scattering amplitudes with energy, causing gravity to become strongly coupled at a scale well below the Planck scale. A prototypical example of a theory obeying the Tower WGC, SDC, and Species Bound is a Kaluza-Klein compactification of a gravitational theory on a circle. The large radius limit of the circle is an infinite-distance limit in which a tower of Kaluza-Klein modes of the graviton appears at the scale $1/R \sim g M_\mathrm{Pl}$, and the higher-dimensional Planck scale supplies the low cutoff $\Lambda_\mathrm{QG}$. In other examples, the tower arises as a set of oscillator modes of a low-tension string; it has been proposed that all weak-coupling limits in quantum gravity follow one of these two paradigms~\cite{Lee:2019wij}.

In summary, the Swampland program has led to a relatively concrete proposal for the nature of weak-coupling limits in quantum gravity. In such cases, we expect to find towers of weakly-coupled states, which drive the theory to strong coupling at a UV scale well below the Planck scale. The minimal violation of global symmetries is parametrically controlled by the separation between the UV cutoff and the Planck scale. These are powerful statements with potential applications in particle phenomenology and cosmology, reviewed in the Snowmass white paper~\cite{whitepaperswamplandpheno}. However, this picture currently rests on surveys of string theory examples and on qualitative arguments. An important goal is to provide either more robust and more quantitative evidence that these are universal statements about quantum gravity, or counterexamples. In the remainder of this section, we will sketch some paths toward those goals, connecting with the other themes of this white paper.

\subsection{Testing the Weak Gravity Conjecture via Positivity Bounds}

Positivity bounds in effective field theory provide a powerful tool for testing strong forms of the WGC. If the Tower WGC is true, then we expect that there are objects of arbitrarily large charge which obey the WGC. For very large charge, we expect these objects to be black holes~\cite{ArkaniHamed:2006dz}. At first glance, this appears to be a contradiction: the WGC is the opposite of the extremality bound obeyed by classical black holes. However, the extremality bound relevant for the WGC is that computed from the two-derivative effective action, which applies to arbitrarily large black holes. A finite size black hole is sensitive to higher derivative corrections, which can allow it to become slightly superextremal. The condition that black holes become superextremal translates into the positivity of a certain linear combination of Wilson coefficients for four-derivative operators~\cite{Kats:2006xp}.  This combination includes operators like $(F_{\mu \nu}^2)^2$ and $(F_{\mu \nu}{\tilde F}^{\mu \nu})^2$, known to have positive coefficients in effective field theory. However, it also includes operators like $R_{\mu \nu \rho \sigma}F^{\mu \nu}F^{\rho \sigma}$, which are more difficult to constrain; and even standard field-theoretic proofs of positivity are in some doubt when the coefficients of higher-dimension operators are Planck-suppressed, due to the difficulty of eliminating the $t$-channel graviton pole in dispersion relation arguments. The connection between the WGC and positivity bounds has led to a great deal of activity investigating how causality, unitarity (including the generalized optical theorem) or other considerations constrain these higher-dimension operators in gravitational EFTs; see, e.g.,~\cite{Cheung:2014ega, Cheung:2018cwt, Hamada:2018dde, Bellazzini:2019xts, Charles:2019qqt,Jones:2019nev, Alberte:2020jsk, Arkani-Hamed:2021ajd, Henriksson:2022oeu}. In four dimensional  Einstein-Maxwell theory, loops generate logarithmic corrections to $F^4$-type operators, which in many cases imply that exponentially large black holes satisfy the WGC~\cite{Charles:2019qqt,Arkani-Hamed:2021ajd}. Another interesting outcome of this line of research is the discovery of a quantitative relationship between corrections to the  black hole extremality bound and corrections to the Wald entropy of the black holes~\cite{Cheung:2018cwt}. In  fact, this entropy/extremality correction can be derived from general thermodynamic considerations~\cite{Goon:2019faz}. In this way, studies of the WGC have had unexpected corollaries for more general investigations of black hole thermodynamics.

\subsection{From Swampland Criteria to Testable QFT Claims via Holography}

While positivity bounds offer a theoretical tool for checking a prediction of the Tower WGC, a more ambitious goal is to directly prove Swampland conjectures from general principles. Although we have limited theoretical tools for studying the full space of quantum gravity theories, AdS/CFT provides a powerful nonperturbative understanding of a subspace. A conjectured property of all quantum gravity theories might be proven, in the asymptotically AdS case, via holography; alternatively, a CFT counterexample might be found. In particular, the statement that quantum gravity theories do not admit global symmetries has been proven, in the asymptotically AdS  context, using the tools of entanglement wedge reconstruction~\cite{Harlow:2018tng}. (CFTs themselves can have global symmetries, but these are gauge, not global, symmetries of the AdS dual.)

While it is encouraging that holography strengthens the argument against global symmetries, it would be very useful to obtain more quantitative results. The use of the conformal bootstrap program has been proposed as a test of the existence of proposed string theory vacua, e.g., the Large Volume Scenario~\cite{Conlon:2018vov}. Swampland criteria can also be translated into conjectures about CFTs; for instance, the WGC translates into an inequality involving the dimension of a charged operator~\cite{Nakayama:2015hga}. More effort is needed to translate the stronger Swampland bounds involving towers of states into the language of CFTs. For some steps in this direction, see~\cite{Alday:2019qrf, Perlmutter:2020buo,Baume:2020dqd}. A difficulty in adapting the Tower WGC to this context is that it refers to the existence of charged single-particle states. In the CFT context, this becomes a  statement about single-trace operators, a notion that only makes sense in CFTs with a large-$N$ expansion. Away from this context, the existence of charged towers of particles might be interpreted as a statement about the growth in the density of states of charged operators of large dimension. We expect that the important scale $\Lambda_\mathrm{QG}$ translates into an operator dimension that is often parametrically smaller than the central charge $c$. For example, in ${\cal N} = 4$ SYM, the 5d Planck scale scales as $c^{1/3} \sim N^{2/3}$, but $\Lambda_\mathrm{QG}$ lies at the 10d Planck scale which is much smaller still, scaling as $c^{1/8} \sim N^{1/4}$. A challenge, then, is to provide rigorous bounds on the density of states of different charge and spin, for operator dimensions in an intermediate regime $1 \ll \Delta \ll c$.

The Charge Convexity Conjecture is an interesting recent attempt to formulate a simple, sharp, Tower WGC-like statement in the CFT context: $\Delta(n q)$, the smallest dimension associated with any operator of charge $n q$, should be a convex function of $q$ (for fixed $n$ of $O(1)$)~\cite{Aharony:2021mpc}. In the context of flat-space quantum gravity, the case $n=1$ is known to fail in the context of certain toroidal orbifold compactifications~\cite{Heidenreich:2016aqi}. For example, one can find a compactification on a smooth $T^3/\mathbb{Z}_2$ geometry for which the states of even charge and odd charge under a particular $U(1)$ gauge group are independently convex, but the full spectrum is not. This suggests that the parameter $n$ in the Charge Convexity Conjecture may depend in an interesting way on additional discrete data associated with the CFT. More generally, discrete data such as  charge assignments play an important role in the application of many Swampland conjectures to phenomenology or cosmology. A general argument that CFTs do not admit low-lying operators of very large charge, for example, could be a very useful input to applications.

\section{Future Directions \& Outlook}

The study of UV constraints on IR physics is a burgeoning field, with many new conceptual and technical developments. It brings together old ideas about the role of causality, locality, and unitarity in quantum field theory; a more modern perspective on EFTs; and modern computational resources that make it possible to map out bounds in high-dimensional parameter spaces. Recent work has brought these developments into close contact with directly experimentally testable ideas, such as SMEFT. The Swampland program has reinvigorated the intersection of quantum gravity with phenomenology, turning attention toward general principles that are testable both theoretically and empirically. It is important to put this circle of ideas on a more rigorous footing, which progress in positivity bounds and the bootstrap program could enable. The future of this field is bright. Here we collect a number of brief remarks on  possible future developments.

\medskip

\noindent {\bf S-matrix bootstrap \& CFT axioms.}-- The CFT setup of section \ref{section:CFT}, from the dual gravity perspective, is probing local high energy scattering deep in AdS. Hence, it is expected that the bounds obtained from the CFT arguments should persist even in the flat space limit. So, it would be important to unify the flat space bounds and the AdS bounds in a systematic way. This can be achieved by rigorously deriving all the assumptions of the S-matrix bootstrap directly from the CFT axioms by taking the flat space limit. Some aspects of this  problem have already been addressed in \cite{Caron-Huot:2021enk}.

\medskip

\noindent {\bf Bootstrapping $\langle TTTT\rangle$.}-- The relationship (\ref{eq:einstein}) is expected to be true even for higher-point stress-tensor correlators in CFTs with large central charge and a sparse spectrum. In particular, it is of importance to impose precise constraints on the 4-pt correlator $\langle TTTT\rangle$, including terms that are suppressed by $\Delta_\mathrm{gap}$. In the dual EFT, this will translate into bounds on graviton 4-pt interactions. One promising approach is to extend the scalar analysis of $\phi^2\Box^k \phi^2$ for the graviton by studying  Regge correlators of the stress tensor operator.

\medskip

\noindent {\bf Higher-point amplitudes \& Positivity Bounds.}-- Almost all discussions of positivity/bootstrap bounds for scattering amplitudes have focused on the case of $2$-$2$ scattering, which gives only limited information on the constraints on the Wilsonian effective action. A major uncharted territory is to extend these bounds following from unitarity, crossing and analyticity to higher-point scattering amplitudes (though see~\cite{Chandrasekaran:2018qmx} for some results in scalar field theory under special conditions). Significant technical challenges remain in the increased complexity of the dependence on multiple Mandelstam variables,  unproven analyticity, and the less transparent positivity conditions. Nevertheless successfully overcoming these challenges would add considerable power and reach to these methods. In particular, constraining operators with higher order in the fields would allow to have a much better handle on SMEFT.  One advantage of the CFT-based approach is that it provides an easier way to study higher-point interactions in AdS. For example, one may consider bootstrapping various CFT four-point correlators of operators that include double-trace (or multi-trace) CFT operators (see \cite{Kundu:2019zsl}).

\medskip

\noindent {\bf SMEFT.}-- The exploration of positivity bounds on the SMEFT parameter space is still in the early stages, and yet we have already seen their constraining power, their usefulness to infer the UV physics and concrete examples of surprising predictions. As the SMEFT contains a large number of dimension-8 operators, it is a formidable task to chart the whole parameter space and quantify the exact boundaries of the bounds to the extent that is usable in experimental data analysis, but nevertheless it is a direction that interests both the theoretical and experimental communities. More importantly, it is crucial to understand how the SMEFT positivity bounds can help with the inverse problem, and particularly how they can sharpen our understanding of the connection between the UV models and the bounds in the presence of many degrees of freedom with internal symmetries and to what extent the UV symmetries can be inferred from the bounds. Another direction needs further investigation is the interplay between the dimension-8 operators and the quadratic terms of the dimension-6 operators.

\medskip

\noindent {\bf Cosmological EFTs.}-- Although as discussed earlier some attempts have been made to connect the S-matrix analyticity methods with cosmological effective theories, as yet the rules are far less clear largely due to the absence of Lorentz symmetry and the associated Mandelstam analyticity. At the same time, cosmological EFTs notably via early universe inflationary theories provide probably our closest link with high energy physics. It is thus of paramount importance to develop a better understanding of how these types of effective theories are constrained by UV/IR considerations.

\medskip

\noindent {\bf From Swampland conjectures to theorems.}-- Many aspects of the Swampland program are well-supported by string theory examples, and in some cases by black hole arguments. Nonetheless, it would be a significant step forward to put them on a firmer footing, via positivity bounds or bootstrap arguments. Many of these conjectures, in the CFT context, characterize the scale $\Delta_\mathrm{gap}$ and the density of states (of various charges and spins) around this scale. Some of  the conjectures may relate to convexity properties of subsets of these  states~\cite{Aharony:2021mpc}. Even more ambitious would be to prove or find convincing counterexamples to a much wider range of conjectures about cosmology (e.g., the existence of approximate de Sitter backgrounds or the behavior of trans-Planckian modes), reviewed in part in~\cite{Brennan:2017rbf, Palti:2019pca, vanBeest:2021lhn, Grana:2021zvf, Harlow:2022gzl}. Because these go beyond asymptotically AdS or asymptotically flat settings, they require further technical developments to make precise statements in time-dependent backgrounds. (See~\cite{Grall:2021xxm} for steps in this direction.)

\medskip

\noindent {\bf Extended operators and higher symmetries.}-- Recently, the physics of $p$-form symmetries, higher group symmetries, and other extended notions of symmetry has received intense study (for a small sample of entry points into this large literature, see~\cite{Aharony:2013hda,Gaiotto:2014kfa,Hofman:2017vwr,Cordova:2019lot,Cordova:2018cvg,Lake:2018dqm,Hofman:2018lfz,McNamara:2019rup, Rudelius:2020orz,Cordova:2020tij, Heidenreich:2021tna}). Aspects of the Swampland program can be expressed in terms of such symmetries; for example, the Weak Gravity Conjecture is closely related to the absence of even approximate 1-form symmetries at the UV cutoff scale~\cite{Cordova:2022rer}. However, rigorous positivity and causality bounds have mostly been applied to correlation functions of local operators, or $S$-matrix elements of particles. Extending these arguments to line and surface  operators could make contact with generalized symmetries and offer new perspectives on the Swampland.

\medskip

Of course, the most important developments are likely those that we cannot yet anticipate. This is already a lively research area within theoretical physics, which will thrive in the presence of strong support for a diverse and ambitious experimental program and for theoretical work that maximizes the information we can extract from experimental data.

%\section{Outlook}

%What are some major open questions? Which areas are progressing most quickly?

~\\
\noindent{\bf Acknowledgements}\\
We thank Grant Remmen and Aninda Sinha for feedback on a draft. SK is supported by the Simons Collaboration Grant on the Non-Perturbative Bootstrap. MR is supported by the DOE Grant DE-SC0013607, the NASA Grant 80NSSC20K0506, and the Alfred P.~Sloan Foundation Grant No.~G-2019-12504. The work of AJT and CdR is supported by STFC grants ST/P000762/1 and ST/T000791/1. CdR thanks the Royal Society for support at ICL through a Wolfson Research Merit Award.  CdR is supported by the European Union Horizon 2020 Research Council grant 724659 MassiveCosmo ERC2016COG, by the Simons Foundation award ID 555326 under the Simons Foundation Origins of the Universe initiative, Cosmology Beyond Einstein's Theory and by a Simons Investigator award 690508. AJT thanks the Royal Society for support at ICL through a Wolfson Research Merit Award. SYZ is supported by National Natural Science Foundation of China under grant No.~11947301, 12075233 and 12047502, and supported by the Fundamental Research Funds for the Central Universities under grant No.~WK2030000036.

\bibliography{ref}
\bibliographystyle{utphys}

\end{document}